\documentclass[a4paper,11pt]{article}
\usepackage{jheppub}
\usepackage[T1]{fontenc}
\usepackage{amsmath,amssymb,amsfonts}%
\usepackage{booktabs,graphicx,array,xspace,wasysym,placeins,hyperref}
\usepackage{multirow}%
\usepackage{mathrsfs}%
\usepackage[title]{appendix}%
\usepackage{xcolor}%
\usepackage{tikz}
\usepackage{cite}
\usepackage[compat=1.1.0]{tikz-feynman}
\usepackage{subcaption}
\usepackage[all]{hypcap}

\newcommand{\MEPSatLO}{M\protect\scalebox{0.8}{E}P\scalebox{0.8}{S}@L\protect\scalebox{0.8}{O}\xspace}

\newcommand{\CSS}{CSS\scalebox{0.8}{HOWER}\xspace}
\newcommand{\Comix}{C\protect\scalebox{0.8}{OMIX}\xspace}
\newcommand{\Ahadic}{A\protect\scalebox{0.8}{HADIC}\xspace}

\newcommand{\Sherpa}{S\scalebox{0.8}{HERPA}\xspace}

\newcommand{\Rivet}{R\scalebox{0.8}{IVET}\xspace}

\newcommand\HERA{H\scalebox{0.8}{ERA}\xspace}

\newcommand{\MSbar}{\ensuremath{\overline{\mathrm{MS}}}}

\hypersetup{hidelinks,
  pdfauthor={Thomas Gehrmann, Peter Meinzinger},
  pdftitle={Photon-induced jet production at future lepton colliders},
  pdfkeywords={QCD, Perturbation Theory}
}

\begin{document}
\preprint{ZU-TH 77/25}
\title{Photon-induced jet production at future lepton colliders}
\author{Thomas Gehrmann and}
\author{Peter Meinzinger}
\affiliation{Physik-Institut, Universit\"at Z\"urich,
  Winterthurerstrasse 190, CH-8057 Z\"urich, Switzerland}
\emailAdd{thomas.gehrmann@uzh.ch}
\emailAdd{peter.meinzinger@uzh.ch}

\abstract{The production of
hadronic final states in electron-positron or electron-hadron collisions is
induced predominantly by quasi-real photons that were emitted off incoming leptons.
In these processes, the photon either enters directly or through its
resolved parton content,
which is at present only loosely constrained by experimental data. We perform
a detailed
phenomenological
study of photon-induced jet production processes in high-energy
$e^+e^-$ collisions, investigating in particular their potential to assess
contributions from the resolved photon structure. }

\keywords{QCD, lepton-lepton collider, photon}

\maketitle

\section{Introduction}

The interaction of
quasi-real photons with other particles at high energies
consists of two components~\cite{Walsh:1973mz,Klasen:2002xb}:
a direct process, where the photon
directly participates in the interaction, and a resolved process,
where the photon splits up into a collinear cluster of particles
prior to the interaction. For hadronic final states, in particular
the primary splitting of the photon into a quark-antiquark pair
and its subsequent partonic evolution are of relevance.
These are described by an inhomogeneous variant~\cite{Witten:1977ju}
of the DGLAP evolution equations~\cite{Altarelli:1977zs,Dokshitzer:1977sg}.

The description of the resolved process thus involves
parton distributions (PDFs) of the photon, which
describe the probabilities of finding partons with
specific momentum fractions $\xi$ at a resolution scale $\mu^2$ inside the photon.
These photon PDFs obey evolution equations~\cite{Witten:1977ju} in $\mu^2$ with
non-perturbative initial conditions. These initial conditions are poorly
constrained by data from $e^+e^-$ and $\gamma p$ collisions,
and are often modelled assuming so-called
vector meson dominance (VMD)
models~\cite{Brodsky:1971vm,Zerwas:1974tf,Gluck:1983mm}
which describe the
content of the photon as a superposition of light vector mesons.

An improved knowledge on the photon PDFs is crucial to
improve the accuracy of predictions for hadronic final states
in $e^+e^-$ and $ep$ collisions
at high energies, which both receive substantial
contributions from initial states with quasi-real photons.
In the light of future high-energy $e^+e^-$ colliders currently
being discussed (FCC~\cite{FCC:2018evy},
CEPC~\cite{CEPCStudyGroup:2018ghi}, ILC~\cite{ILC:2013jhg})
and of the upcoming BNL
Electron-Ion Collider (EIC,~\cite{Accardi:2012qut}), these predictions will become
crucial both for physics studies and for background predictions
to other processes under consideration that yield hadronic final
states.

Currently available parametrizations~\cite{Gluck:1991jc,Abramowicz:1991yb,Gordon:1991tk,Schuler:1995fk,Gluck:1999ub}
of the photon PDFs rely on fits to
$e^+e^-$ and $\gamma p$ data in combination with VMD models.
Lacking recent data on these processes, no updates or
refinements on the photon PDFs have been performed for nearly
three decades. An overview on the status of data and theory 
relevant to photon PDFs can be found in~\cite{Krawczyk:2000mf}. 

In this article, we investigate the reach and
sensitivity of future measurements on hadronic final states
resulting from $e\gamma$-initiated processes in $e^+e^-$ collisions,
in particular in view of probing the resolved photon
contributions.
In Section~\ref{sec:lepton-photon}, we describe the simulation framework used in our
study. The main results are discussed in Section~\ref{sec:results}, followed by a
summary of our main findings
in Section~\ref{sec:conc}.

\section{Setup for photon-lepton scattering}\label{sec:lepton-photon}

We consider $e^-e^+$ beams with beam energies of 125 GeV using a
pre-release version of \Sherpa 3.1~\cite{Sherpa:2024mfk} for
the simulation of the process
\begin{equation}
    e^- (p_1) + e^+(p_2) \xrightarrow{\to e^- \gamma} e^-(p_1^\prime) + e^+(p_2^\prime) + X
\end{equation}
where $X$ denotes any hadronic final state.
The process is considered analogously to deep inelastic scattering (DIS),
i.e.\ one lepton emits a quasi-collinear photon while the other one scatters at a large angle with that photon.
The flux of quasi-real photons is calculated in the Weizs\"acker-Williams approximation~\cite{vonWeizsacker:1934nji,Williams:1934ad,Budnev:1975poe}
\begin{align}
    f_{\gamma / e^-}(z) = \frac{\alpha_\mathrm{em}}{2 \pi} \frac{\mathrm{d}z}{z}
                \left[ \left( 1 + (1 - z)^2 \right) \log \left( \frac{Q^2_\mathrm{max}}{Q^2_\mathrm{min}} \right) - 2 m_e^2 z^2 \left( \frac{1}{Q^2_\mathrm{min}} - \frac{1}{Q^2_\mathrm{max}} \right) \right]
\end{align}
with an assumed maximum virtuality of $Q^2_\mathrm{max} = 4 \ \mathrm{GeV}^2$ and with the minimum virtuality given by $Q^2_\mathrm{min} = \frac{m_e^2 z^2}{1-z}$.

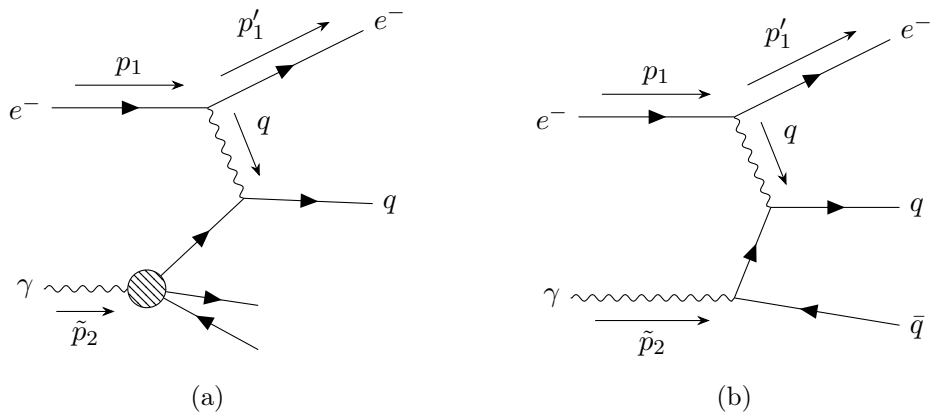
\begin{figure*}[t]
    \centering
    \begin{subfigure}[t]{0.45\textwidth}
        \centering
        \begin{tikzpicture}[scale=0.8,align=center,node distance=1cm]
        \begin{feynman}
            \vertex (i1)  at (-3,  1) {\(e^-\)};   
            \vertex (v1)  at (0, 1);  
            \vertex (o1)  at ( 3,  2.5) {\(e^-\)};  
            \vertex (v2)  at ( 0.60,  -0.5); 
            \vertex (g)   at (-3, -2)  {\(\gamma\)};  
            \vertex [small,blob] (B) at (-1,-2) {}; 
            \vertex (j1)  at ( 3,  -0.60)  {\(q\)}; 
            \vertex (r1)  at ( 1, -2.3)  {}; 
            \vertex (r2)  at ( 1, -3.1)  {}; 

            \diagram*{
            (i1) -- [fermion, momentum=\(p_1\)] (v1)
                    -- [fermion, momentum=\(p_1^\prime\)] (o1),

            (v1) -- [photon, momentum=\(q\)] (v2),

            (g)  -- [photon, momentum'=\(\tilde{p}_2\)] (B),

            (B)  -- [fermion] (v2),

            (v2) -- [fermion] (j1),

            (B)  -- [fermion] (r1),
            (B)  -- [anti fermion] (r2),
            };
        \end{feynman}
        \end{tikzpicture}
        \caption*{(a)}
    \end{subfigure}
    \begin{subfigure}[t]{0.45\textwidth}
        \centering
        \begin{tikzpicture}[scale=0.8,align=center,node distance=1cm]
        \begin{feynman}
            \vertex (i1)  at (-3,  1) {\(e^-\)};   
            \vertex (v1)  at (0, 1);  
            \vertex (o1)  at ( 3,  2.5) {\(e^-\)};  
            \vertex (v2)  at ( 0.60,  -0.5); 
            \vertex (g)   at (-3, -2)  {\(\gamma\)};  
            \vertex (v3) at (0,-2); 
            \vertex (j1)  at ( 3,  -0.50)  {\(q\)}; 
            \vertex (j2)  at ( 3, -2.5)  {\(\bar{q}\)}; 

            \diagram*{
            (i1) -- [fermion, momentum=\(p_1\)] (v1)
                    -- [fermion, momentum=\(p_1^\prime\)] (o1),

            (v1) -- [photon, momentum=\(q\)] (v2),

            (g)  -- [photon, momentum'=\(\tilde{p}_2\)] (v3),

            (v3)  -- [fermion] (v2),

            (v2) -- [fermion] (j1),

            (v3)  -- [anti fermion] (j2),
            };
        \end{feynman}
        \end{tikzpicture}
        \caption*{(b)}
    \end{subfigure}
\caption{Examples of diagrams contributing to deep-inelastic scattering of an electron on a collinear quasi-real photon.
(a) The photon can be resolved and interact hadron-like.
(b) The direct contribution contributes at the level of two out-going partons. }
\label{fig:lepton-photon-scattering}
\end{figure*}

The total cross-section for jet production through quasi-real photons is then
the sum of two types of contributions:
(i) direct contribution, i.e.\ $e \gamma \to e j j$, and
(ii) the resolved contribution where the photon behaves hadron-like by resolving a parton $p$, i.e.\ $e p \to e j$.
We show exemplary diagrams for both contributions in Fig.~\ref{fig:lepton-photon-scattering}.
In our computation, the hadron-like component is given by the SAS1M photon PDF~\cite{Schuler:1995fk} which employs the \MSbar scheme. 
At higher orders in perturbation theory, mass factorization and parton 
evolution~\cite{Witten:1977ju} result in a mixing between
both contributions, which can however still be assigned uniquely according to the underlying parton-level subprocesses. 

Summing over the direct- and resolved-photon contributions, one has to make sure to coherently add these.
While at lowest order, the resolved photon contributes at the level of $e^- p \to e^- j$,
the direct photon only enters at the next multiplicity, i.e.\ $e^- \gamma \to e^- j j$, at the matrix-element level.

This issue can be overcome by employing merging to coherently add
resolved and direct channels. Multi-leg merging  creates inclusive samples from matrix-elements of different multiplicities while still
allowing for resummation through parton showers in the relevant regions of phase space.
This is achieved by phase-space slicing using a jet criterion that separates soft and hard regions,
where the latter are described by matrix elements and the former by the parton shower.
Here we adopt the \MEPSatLO prescription~\cite{Catani:2001cc,Hoeche:2009rj}.

In this study, the merging of the resolved (pure QCD) channels is
straightforward and follows previous studies~\cite{Carli:2010cg}.
To merge the direct channels, however, we have to consider a democratic QCD+QED evolution
to allow for a clustering of the $\gamma \to q \bar{q}$ vertex~\cite{Hoeche:2009xc}.
Additionally, we have to enforce a clustering down to a $2 \to 2$ process
such that the available phase space is filled and a coherent addition to the resolved channels is achieved.
These contributions amount to the inclusion of the so-called anomalous components
of the photon, which are part of the PDF evolution encapsulated in terms
proportional to the $P_{\gamma q}$ DGLAP splitting kernel.

We calculate events with up to three jets at LO,
\begin{align}
    e^- e^+ &\xrightarrow{\to e^-\gamma} e^- j j + 0,1j \quad \oplus \nonumber \\
    e^- e^+ &\xrightarrow{\to e^-\gamma \to e^-p} e^- j + 0,1,2j
\end{align}
and the merging scale $Q_{\mathrm{cut}}$ is set dynamically depending on the virtuality~\cite{Carli:2010cg}, as
\begin{equation}
  Q_{\mathrm{cut}} = \frac{\bar{Q}_{\mathrm{cut}}}{\sqrt{1+\frac{\bar{Q}^2_{\mathrm{cut}}}{S_{\mathrm{DIS}} Q^2}}}
  \label{eq:Qcut}
\end{equation}
with $\bar{Q}_{\mathrm{cut}} = 10$ GeV and $S_\mathrm{DIS} = 1$.
The factorisation and renormalisation scale are set according to
$\mu_R^2 = \mu_F^2 = Q^2/4$ for $2\to 2$ core processes and
$\mu_R^2 = \mu_F^2 = (Q^2 + H_T^2)/4$ for higher multiplicities,
where $H_T$ is the scalar sum of the transverse momentum of partons in the final state, evaluated in the laboratory frame.

The strong coupling is set to $\alpha_S(M_Z) = 0.118$ with 4-loop running and electroweak parameters are set using the $G_\mu$ scheme.
We compute tree-level matrix elements with \Comix~\cite{Gleisberg:2008fv} and utilize \CSS~\cite{Schumann:2007mg} and \Ahadic~\cite{Chahal:2022rid} for parton-showering and hadron fragmentation, respectively.
The factorisation and renormalisation scales are varied by factors of 2 to estimate missing higher-order corrections.

Events are then analysed using an interface to \Rivet~\cite{Bierlich:2024vqo} by imposing a cut on both the virtuality $Q^2$ and
the inelasticity $y$ as computed with respect to the scattered electron,
which is determined as the electron with the highest transverse energy.
The computation follows in analogy to HERA analyses with
the small-angle outgoing lepton taking the role of the hadron.
In our setup, we choose the scattered lepton to be the electron, travelling in
the $+z$ direction, and the photon being emitted by the positron, going in the $-z$ direction.
The results generalize trivially to the opposite case.

The assumed detector acceptance is $\left| \eta \right| < 3$ and events are required
to fulfill $Q^2 > 16$~GeV$^2$ and $0.1 < y < 0.9$, based on similar cuts from DIS analyses at HERA.
Unless stated explicitly, we do also apply the
$\left| \eta \right| < 3$  cut on the forward rapidity of the out-going lepton.
Jets are clustered with the anti-$k_T$ algorithm with radius $R =1.0$ and a requirement of $p_T > 5$ GeV. Variations of $0.4<R<1.0$ 
do not affect our findings below.

\section{Results}\label{sec:results}
The kinematical reach of the measurement is illustrated by the distributions in
virtuality $Q^2$ and $x_\gamma$.
These are defined as
\begin{equation}
    Q^2 = -q^2 \quad \mathrm{and} \quad x_\gamma = \frac{Q^2}{Q^2 + W^2} \ .
\end{equation}
We additionally define the momentum ratio with respect to the incoming lepton, $x$, in analogy to the Bjorken variable, and the ratio $z$ as
\begin{equation}
    x = \frac{-q^2}{2 q \cdot p_2} \quad \mathrm{and} \quad z = \frac{x}{x_\gamma} \ .
\end{equation}
It should be noted that both $x$ and $x_\gamma$ use definitions
which in normal DIS settings would be equivalent,
but in this context yield different interpretations.
The definition of $x$ is with respect
to the incoming positron that radiates the quasi-real photon.
The definition of $x_\gamma$, however, uses the mass of the hadronic final state, $W$,
and is hence a ratio with respect to the underlying $e\gamma$ kinematics.
Therefore, the ratio $z$ can be interpreted as the momentum fraction of the lepton taken by the photon.

We first study the sensitivity to this process by considering the impact of the
forward acceptance of the detector and by separating the direct and resolved components.
Even though the distinction into these two components cannot be maintained strictly at higher orders,
it is still informative and allows conclusions on the resulting impact on photon  PDF fits.

\begin{figure*}[t]
\centering
\begin{tabular}{cc}
    \includegraphics[width=0.49\linewidth]{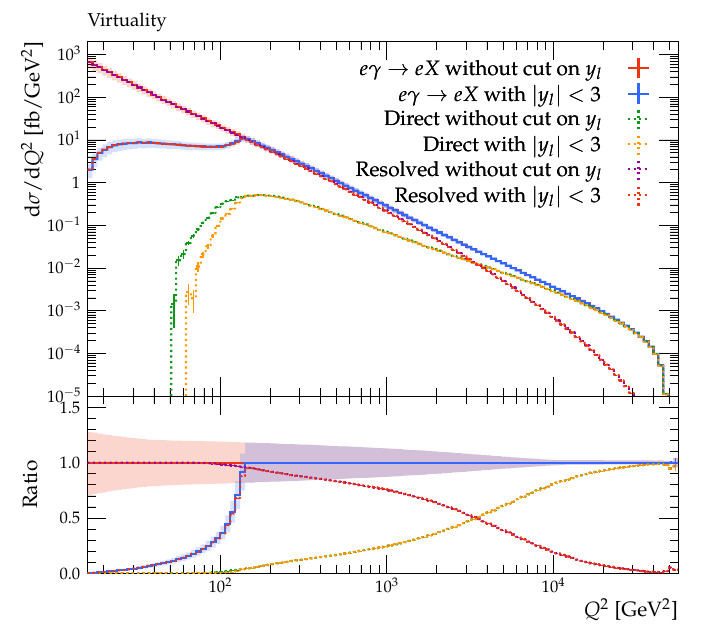} &
    \includegraphics[width=0.49\linewidth]{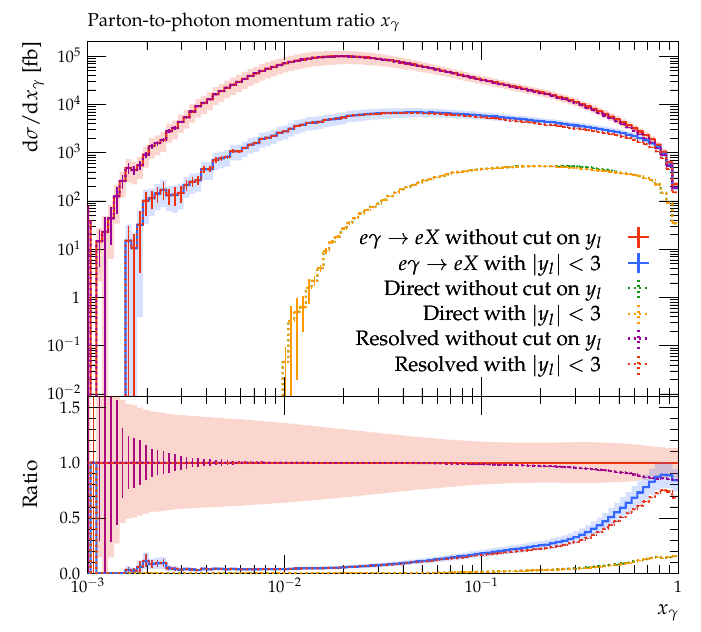} \\
    \includegraphics[width=0.49\linewidth]{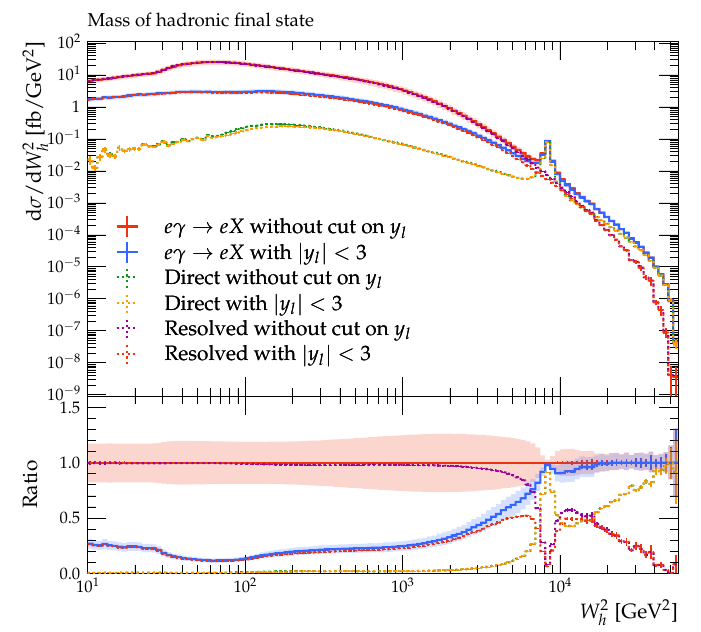} &
    \includegraphics[width=0.49\linewidth]{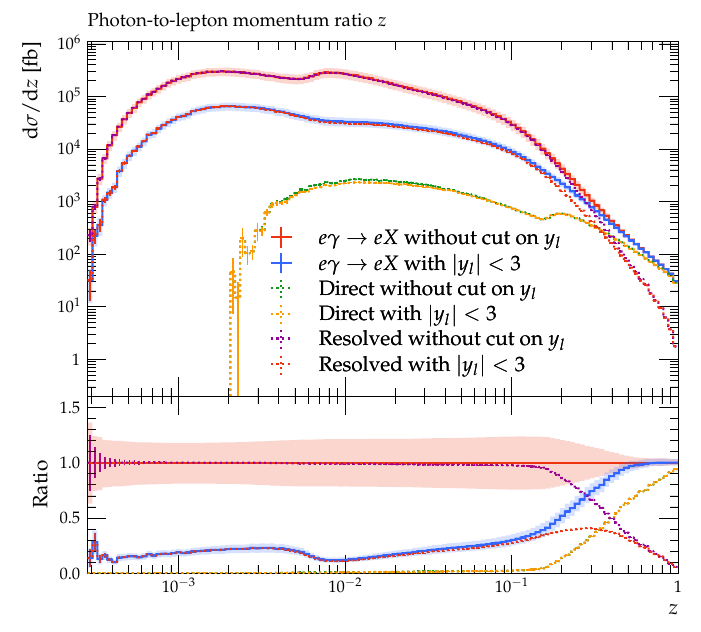}
\end{tabular}
\caption{Distributions in virtuality $Q^2$ (top left), parton-to-photon momentum ratio $x_\gamma$ (top right),
hadronic mass $W^2_\mathrm{h}$ (bottom left) and photon-to-lepton momentum ratio $z$ (bottom right)
for $e \gamma \to e X$, with and without cut on the rapidity of the out-going lepton,
and separating the direct and resolved contributions to the cross-section.}\label{fig:components}
\end{figure*}

In Fig.~\ref{fig:components} it can be seen that indeed the resolved component
is dominating the total cross-section, underlining the potential to fit the
non-perturbative parton content of the photon.
However, the forward acceptance of the scattered lepton is the limiting factor
in the small-$Q^2$ and small-$x_\gamma$ region, severely constraining the availability of data in that region.
The lepton rapidity cut sets in quite sharply at around $Q^2\simeq 12$~GeV$^2$ and $x_\gamma\simeq 0.8$, and leads to a suppression of
the distributions by 90\% in $x_\gamma$ and by 99\% in $Q^2$.
Towards larger values of $Q^2$ and $x_\gamma$, the contribution from
the direct component increases gradually, becoming
dominant in the region $Q^2 > 4 \cdot 10^3$ GeV$^2$ and up to 15\% in the limit $x_\gamma \to 1$. The distribution of the mass of the hadronic final state is again largely dominated by the resolved component. However, it distinctly shows a $Z^0$ peak in the direct channels, above which these channels dominate. This peak is also visible in the $z$ distribution, which marks the threshold for the $\gamma e \to Z^0e \to jje$ configuration. In terms of forward acceptance, we also observe a depletion of the signal for values of $W^2< 10^4$~GeV$^2$ and $z<0.4$ to about 20\%.

\begin{figure*}[t]
\centering
\begin{tabular}{c}
    \includegraphics[width=0.49\linewidth]{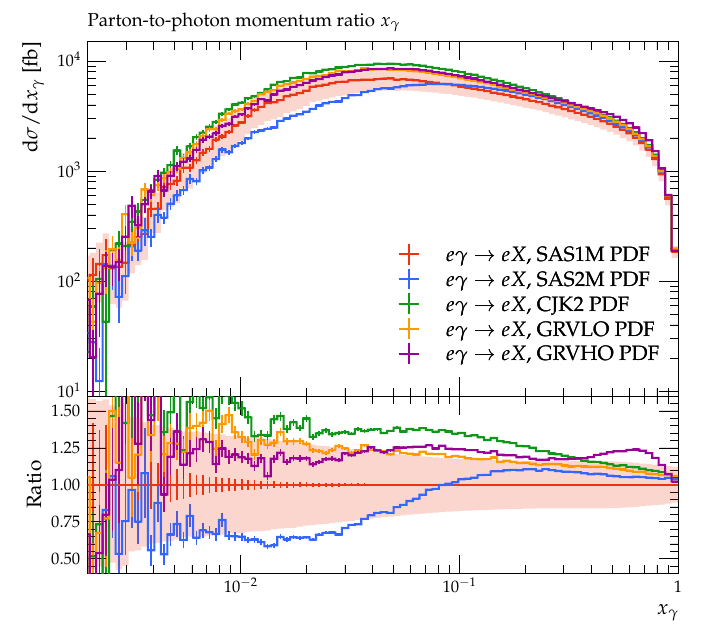}
\end{tabular}
\caption{Distributions in parton-to-photon momentum ratio $x_\gamma$ for different photon PDF parametrisations.}\label{fig:pdf-comparison}
\end{figure*}

Future lepton collider measurements of these distributions 
will enable improved determinations of the photon PDFs. 
To illustrate the sensitivity on the photon PDF, 
Fig.~\ref{fig:pdf-comparison} compares the 
$x_\gamma$ distributions for different photon PDF parametrizations.
Alongside the SAS1M set used throughout this study, predictions for four other photon PDF sets are used:
the SAS2M set, fitted to the same data as SAS1M but with different input parameters,
the GRVLO~\cite{Gluck:1991jc} and GRVHO sets~\cite{Gluck:1991ee}, which were widely used at \HERA,
and lastly the CJK2 set~\cite{Cornet:2004ng}, which is one of the most recent fits.
The latter three fits use the DIS$_\gamma$ factorisation scheme and GRVHO is the only NLO set (substantially 
differing from the other sets in particular at large $x_\gamma$ due
to the NLO correction to the inhomogeneous term in the PDF evolution).
For these parametrizations, we observe 
substantial deviations in the $x_\gamma$ distribution, up to 40\% with respect to the baseline SAS1M fit.
Notably, the small-$x_\gamma$ region is poorly constrained and the differences between
 parametrizations exceed the size of the 
scale variations. 
Future lepton-collider measurements of 
hadronic final states in electron-photon scattering 
will thus provide important new constraints on future 
photon PDF determinations.

\begin{figure*}[t]
\centering
\begin{tabular}{cc}
    \includegraphics[width=0.49\linewidth]{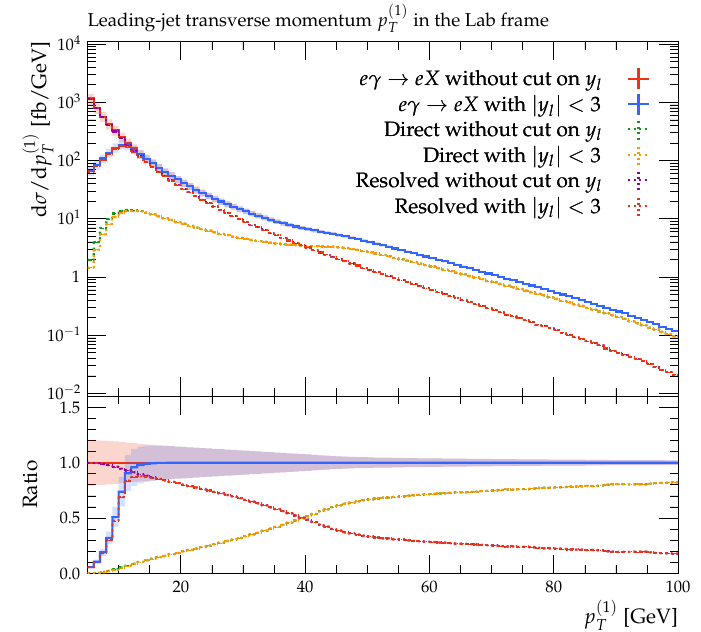} &
    \includegraphics[width=0.49\linewidth]{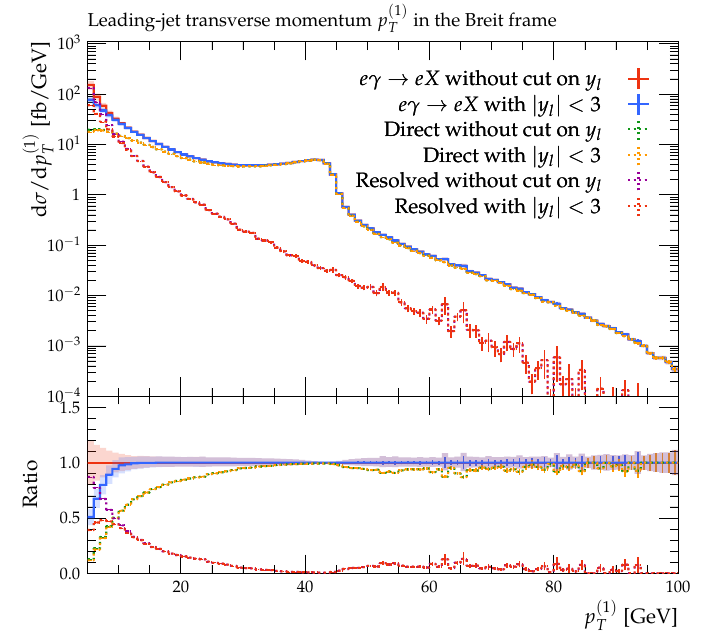} \\
    \includegraphics[width=0.49\linewidth]{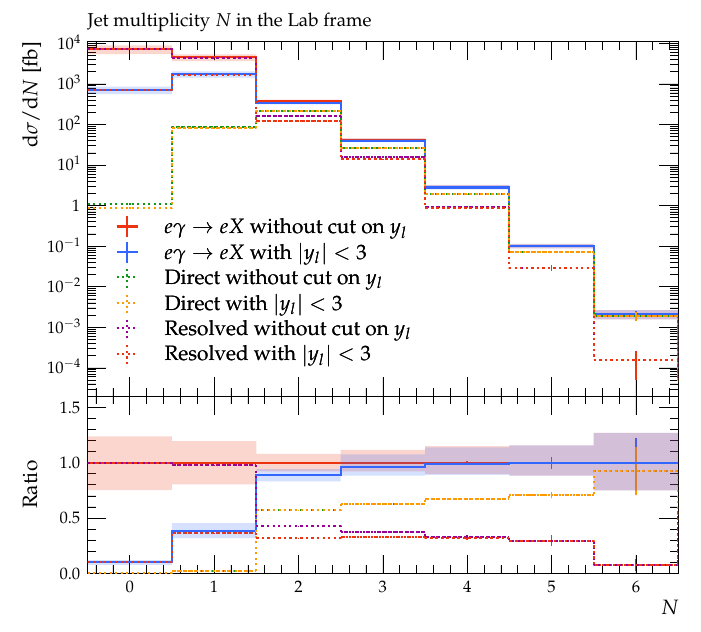} &
    \includegraphics[width=0.49\linewidth]{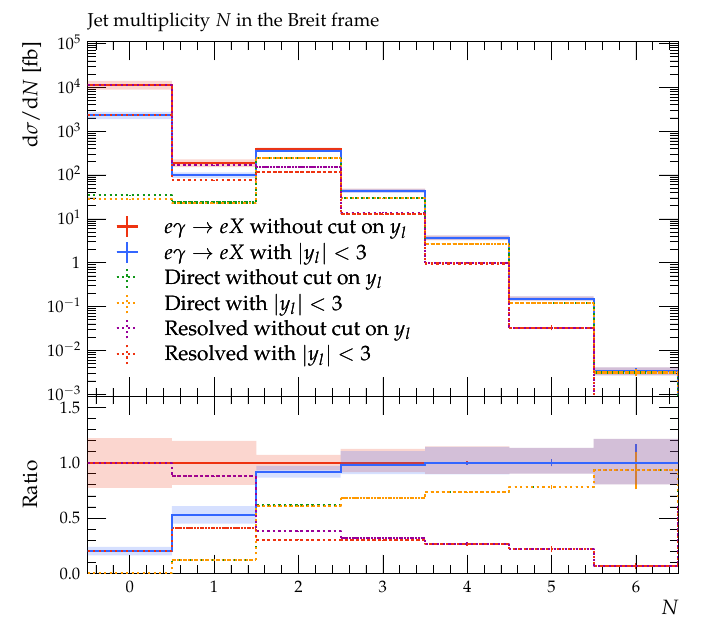}
\end{tabular}
\caption{Distributions in leading jet transverse momentum $p_T^{(1)}$ (top row) and
jet multiplicity $N$ (bottom row) in the laboratory (left column) and
Breit frame (right column) for $e \gamma \to e X$, with and without cut on the
rapidity of the out-going lepton, and separating
the direct and resolved contributions to the cross-section.}\label{fig:jet-obs}
\end{figure*}

We present leading-jet transverse momentum and jet multiplicity as
complementary observables in Fig.~\ref{fig:jet-obs},
displayed for both the laboratory frame and the Breit frame which is
usually chosen for analyses of deep-inelastic scattering.
It is defined as the frame in which the virtual photon acquires a
four-momentum $q = (0, 0, 0, Q)$ and
it can hence be fully determined by measuring the outgoing lepton.
Its advantage is the natural separation into a hadronic and a leptonic hemisphere of the event.

In the Breit frame, the direct component is enhanced over the resolved one.
This is a consequence of the fact that the resolved Born-level scattering process $eq\to eq$
results in a jet at finite transverse momentum only in the laboratory frame.
Any jets in the Breit frame are a result of additional emissions,
either from the parton-shower or from higher-multiplicity corrections in the hard  matrix element.
In contrast, the direct Born-level scattering process $e\gamma \to e q\bar q$
yields two-jet final states both in the laboratory and the Breit frame.

Consequently, the $N=1$ jet fraction is lower by almost two orders of magnitude in the
Breit frame (where it is comparable to the
$N=2$ jet fraction) than in the laboratory frame, and the normalization of the leading-jet transverse momentum
distribution is suppressed by a similar factor. Moreover, the resolved component yields a
sizable contribution in the Breit frame only for low jet transverse momenta $p_T < 10$ GeV.
The forward rapidity cut affects the $p_T$ distributions
below 10 GeV and the multiplicity for $N \le 1$ in both frames.
From these observations, we conclude that  jet cross sections in the Breit frame are not suitable for a study of the
non-perturbative component of photon PDF, where the focus is on the resolved component.

Turning to jet clustering in the laboratory frame in the left-hand column in Fig.~\ref{fig:jet-obs},
we see that 0- and 1-jet configurations dominate.
In fact, the latter yields the highest cross-section once the forward acceptance of the lepton is taken into account.
Comparing resolved and direct contributions, we can see that the latter starts to
dominate in regions of
high jet transverse momenta, $p_T > 40$ GeV, and high multiplicities, $N > 2$.

\begin{figure*}[t]
\centering
\begin{tabular}{cc}
    \includegraphics[width=0.49\linewidth]{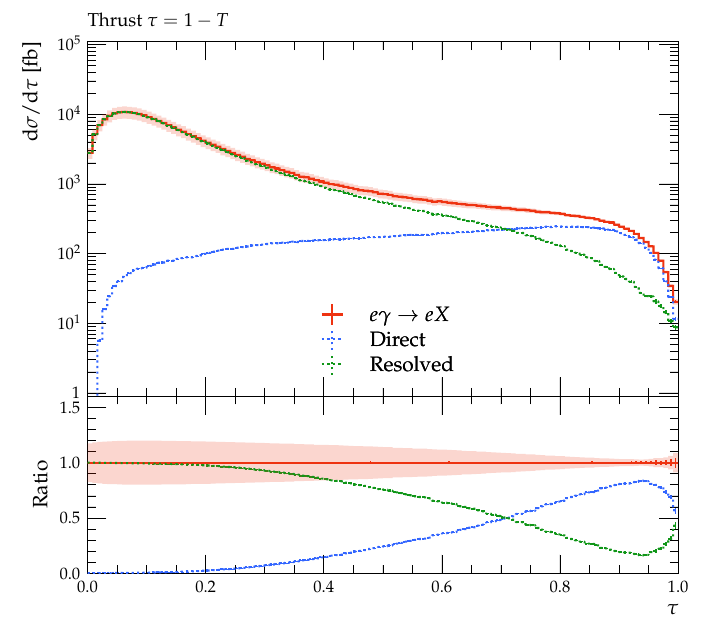} &
    \includegraphics[width=0.49\linewidth]{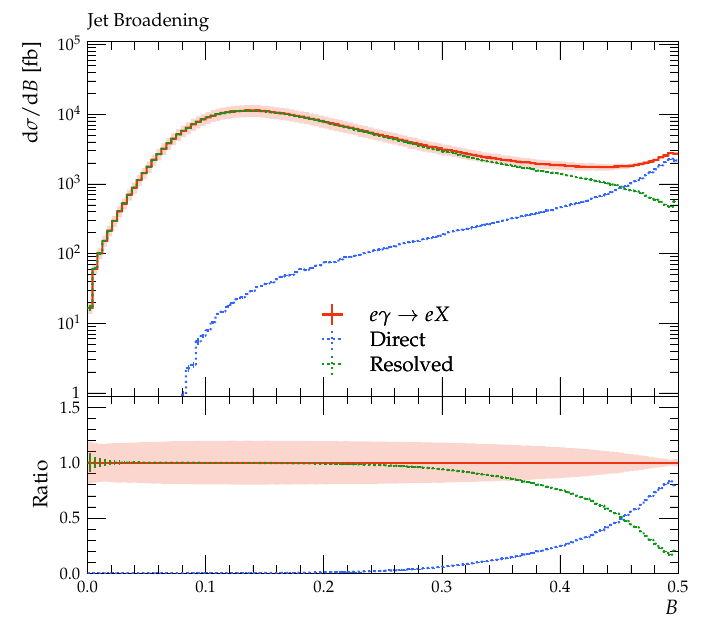}
\end{tabular}
\caption{Event shape distributions
in the variables thrust $\tau = 1 - T$ (left) and jet
broadening $B$ (right)
for $e \gamma \to e X$, separating the direct and resolved contributions to the cross-section.}\label{fig:event-shapes}
\end{figure*}

Event shape distributions provide an alternative to jet cross sections, offering access to
a variety of hadronic final state characteristics. In the Breit frame in lepton-hadron scattering,
event shape variables can be defined in close analogy to electron-positron annihilation. Consequently,
event shape studies are usually performed in the Breit frame.

We focus on two commonly used event shape variables, thrust
and jet broadening, which are both defined
with respect to the axis of the exchanged boson as
\begin{equation} 
    \tau = 1 - T = 1 - \frac{\sum_h \left| p_{z,h} \right|}{\sum_h \left| \vec{p}_{h} \right|} \quad
    \ \mathrm{and} \quad \
    B = \frac{\sum_h \left| p_{\perp,h} \right|}{2 \sum_h \left| \vec{p}_{h} \right|} \ .
\end{equation}
All momenta are evaluated in the Breit frame and the definitions follow an analysis by the H1
collaboration~\cite{H1:2005zsk} of event shapes in electron-proton collisions at DESY HERA.
By defining event shape variables relative to the boson axis, the event-shape distributions favour configurations
where all hadronic activity is collimated along the $z$-axis in the Breit frame (as opposed to a principal
back-to-back axis in electron-positron annihilation), thereby probing information that is highly complementary
to jet distributions in the Breit frame, which require objects with larger transverse momenta.

This feature can be seen very clearly in the thrust and jet broadening
distributions
in Fig.~\ref{fig:event-shapes}.
For the bulk of the phase space, the resolved channels yield the dominant contribution to the
event shape distributions. The direct component yields significant contributions only towards the
kinematical edges $\tau \to 1$ or $B\to 0.5$, which correspond to events
with large transverse momenta in the Breit frame.
Thrust and jet broadening may thus be well-suited observables to separate
resolved and direct contributions in $e\gamma$ collisions, suggesting  a boundary around
$\tau \approx 0.72$ and $B \approx 0.45$.

\begin{figure*}[t]
\centering
\begin{tabular}{cc}
    \includegraphics[width=0.49\linewidth]{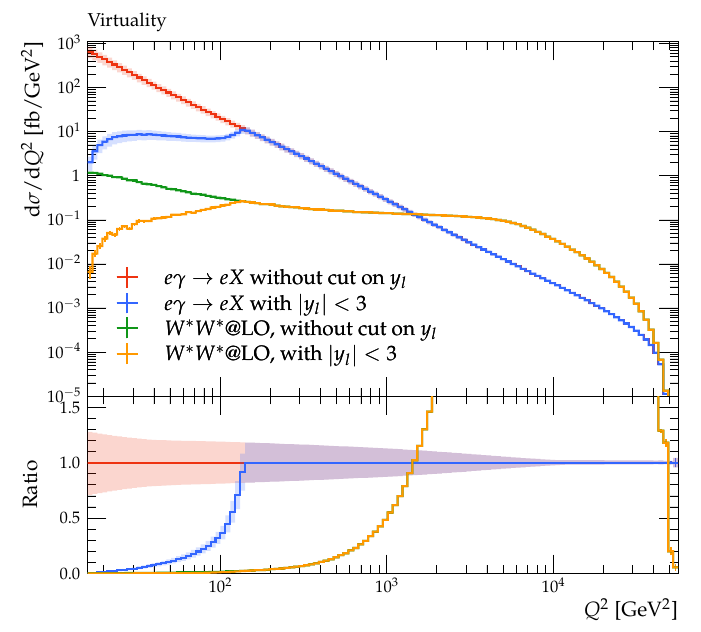} &
    \includegraphics[width=0.49\linewidth]{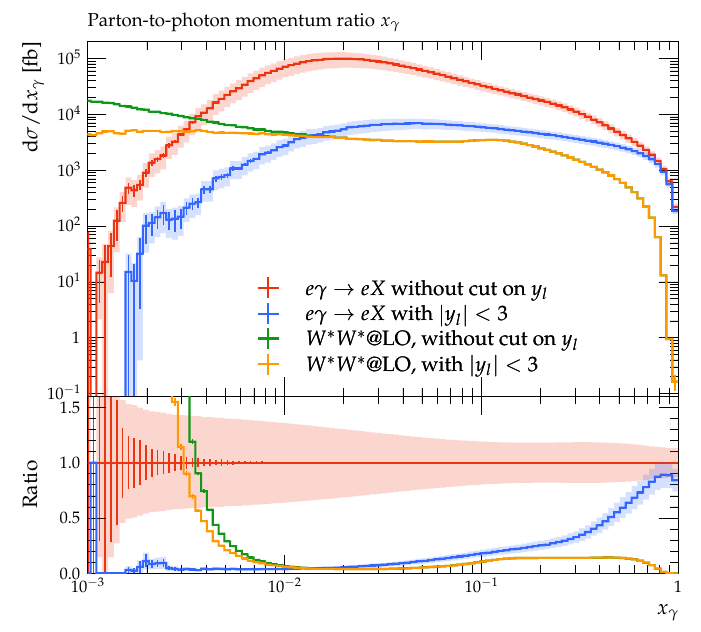}
\end{tabular}
\caption{Distributions in virtuality $Q^2$ (left) and Bjorken variable $x_\gamma$ (right),
for the  $e \gamma \to e X$ and $e^+e^- \to W^\ast_{\to l \bar{\nu}} W^\ast_{\to jj}$
processes,
with and without cut on the rapidity of the out-going lepton.}\label{fig:background-wo-cut}
\end{figure*}

\begin{figure*}[t]
\centering
\begin{tabular}{cc}
    \includegraphics[width=0.49\linewidth]{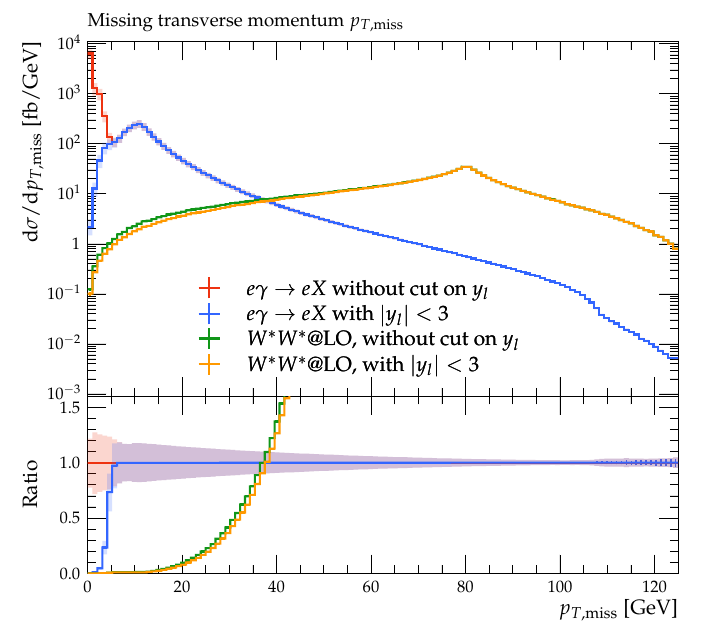} &
    \includegraphics[width=0.49\linewidth]{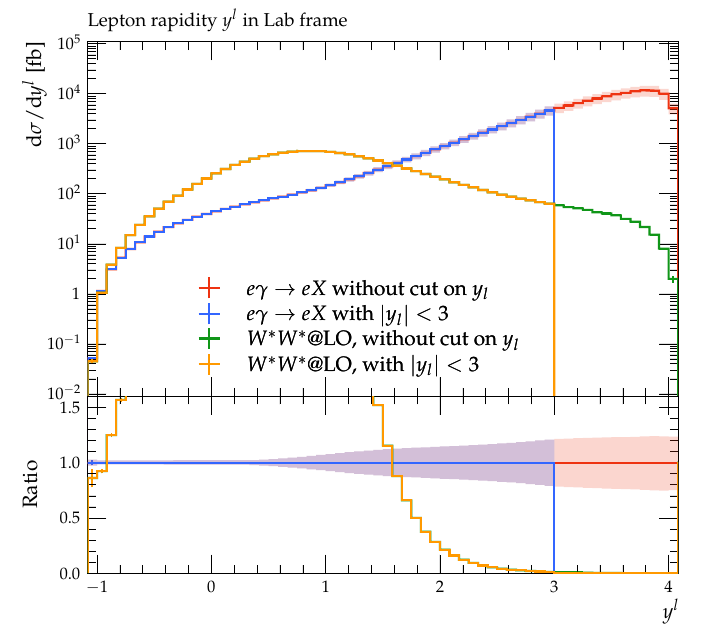}\\
    \includegraphics[width=0.49\linewidth]{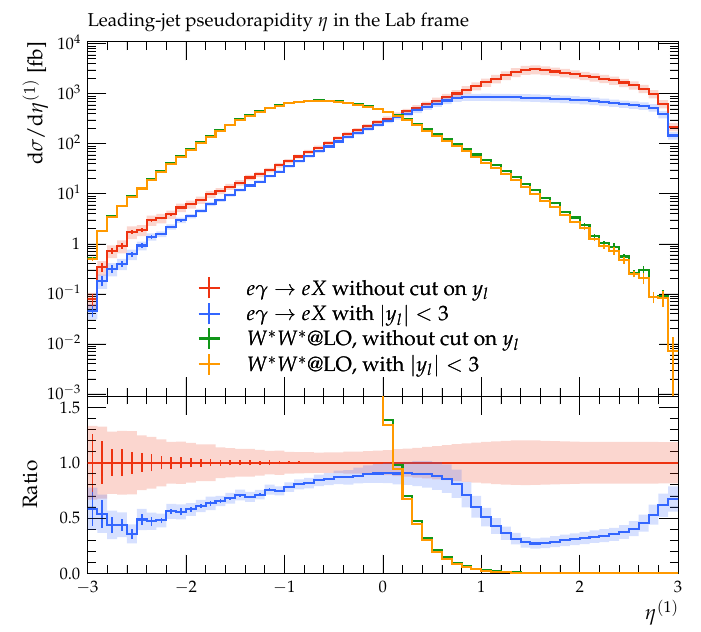} &
    \includegraphics[width=0.49\linewidth]{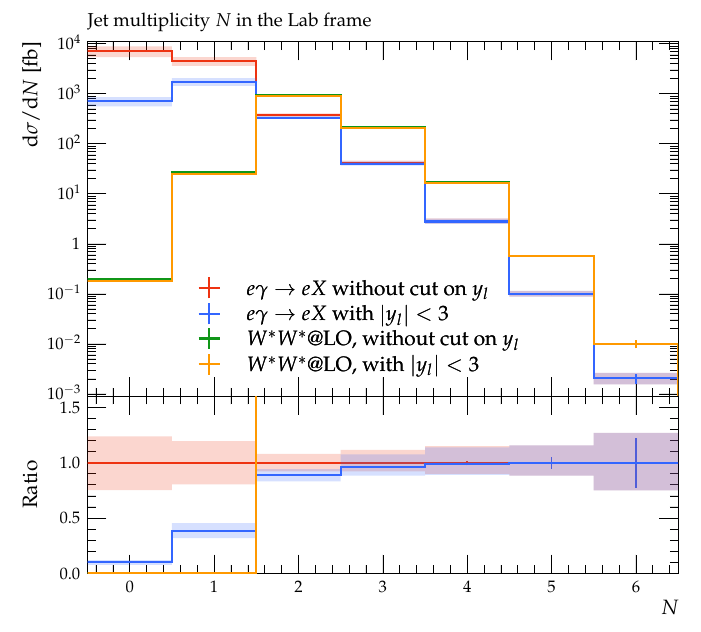} \\
\end{tabular}
\caption{Distributions in
missing transverse momentum $p_{T,\mathrm{miss}}$ (top left), lepton rapidity $y_l$ (top right),
leading-jet pseudorapidity $\eta^{(1)}$ (bottom left) and
jet multiplicity $N$ (bottom right) in the laboratory frame
for the  $e \gamma \to e X$ and $e^+e^- \to W^\ast_{\to l \bar{\nu}} W^\ast_{\to jj}$
processes,
with and without cut on the rapidity of the out-going lepton.}\label{fig:background-dist}
\end{figure*}

The final-state signature of the $e\gamma\to e+$hadrons process
at $e^+e^-$ colliders consists of a single identified
lepton
plus
final-state hadronic activity. A similar final-state signature
can be produced through $W$-boson pair production
$e^+e^- \to W^\ast W^\ast \to l \bar{\nu} jj$ as the neutrino is not detected.
To arrive at predictions for this background, we again use \Sherpa with an
analogous setup as described in Sec.~\ref{sec:lepton-photon},
but calculating $e^- e^+ \to e^- \bar{\nu} j j$ at leading order.

In Fig.~\ref{fig:background-wo-cut}, we compare the two processes with and without cuts on the forward acceptance of the lepton.
We observe that the background of $W$ pair production dominates the event rates for high values of $Q^2$,
starting at around $Q^2 \simeq 1.5 \cdot 10^3$ GeV$^2$.
In $x_\gamma$, the $W$ pair background is negligible for $x_\gamma > 4 \cdot 10^{-3}$ as long as forward leptons are detected.
Once a cut on this acceptance is included, the background contributes a
significant fraction to the cross-section for $x_\gamma \lesssim 0.1$ and then
becoming dominant for values of $x_\gamma < 10^{-2}$.

To identify potential ways of discrimination between the two
processes, we investigate several kinematical distributions in
Fig.~\ref{fig:background-dist}:
missing transverse momentum $p_{T,\mathrm{miss}}$,
lepton  rapidity $y_l$,
leading-jet pseudorapidity $\eta^{(1)}$ and
jet multiplicity $N$ in the laboratory frame.
We observe that the differential cross-section of the lepton-photon
scattering increases towards higher rapidities of the
scattered lepton,
whereas the $W$ pair production drops off in that region of phase space.
A cut on the lepton rapidity due to limited
forward detector acceptance will thus mainly diminish the
signal process, while leaving the background largely unaffected.
The missing transverse momentum peaks as expected at around the $W$ mass in the $W$ pair production
and at small values for the lepton-photon scattering, making this
variable a potential separator between
signal and background.

The pseudorapidity distribution of the leading jet is remarkably
different for both processes. Here, we recall that
the forward rapidity direction is determined by the electron beam, and that only
final states with identified electrons are considered. The
$\eta^{(1)}$ distribution is peaked in the forward region, corresponding to the direction of the incoming electron, while the
$W$-pair production yields final states with leading jets preferably in the backward direction, resulting from the charge and spin
correlations in the underlying Born process.
Finally,
another potential discriminator is the jet multiplicity in the laboratory frame,
where the $e\gamma\to$hadrons yields predominantly zero-jet and
one-jet final
states, while $W$-boson pair production in the semi-leptonic
decay mode produces two-jet final states.

\begin{figure*}[t]
\centering
\begin{tabular}{ccc}
    \includegraphics[width=0.49\linewidth]{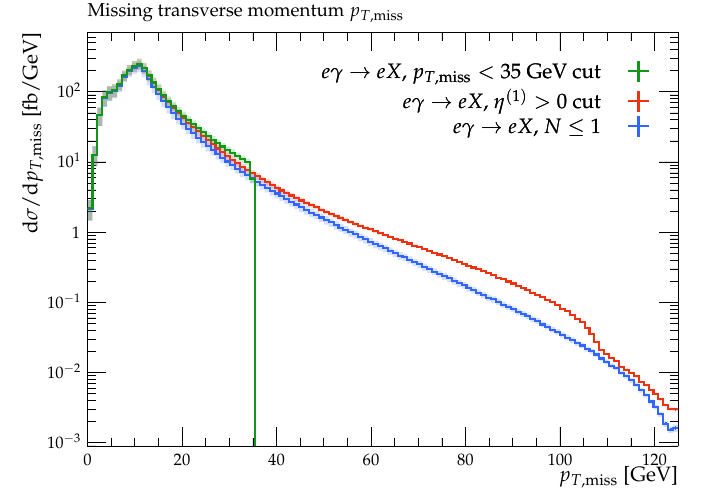} &
    \includegraphics[width=0.49\linewidth]{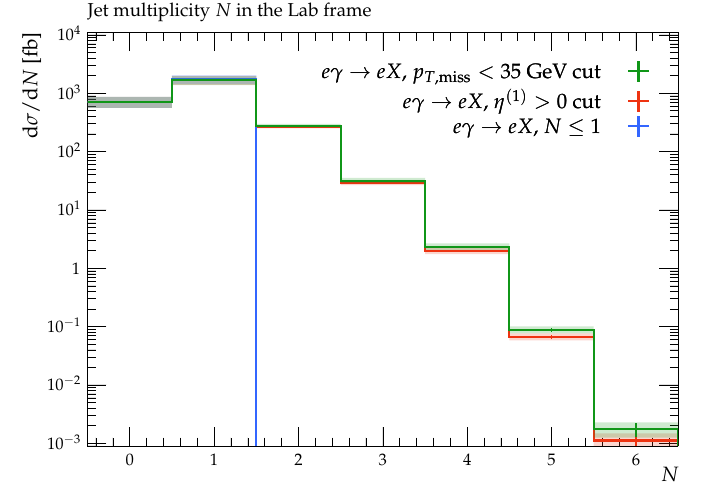} \\
    \includegraphics[width=0.49\linewidth]{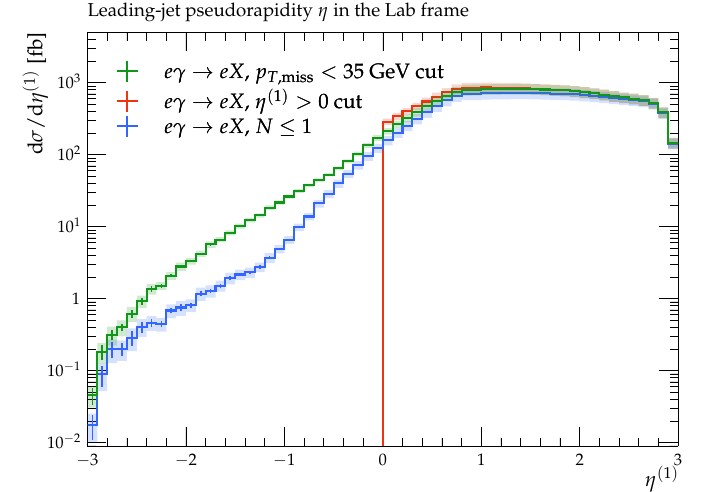}
\end{tabular}
\caption{Distributions of missing transverse momentum $p_{T,\mathrm{miss}}$ (left),
jet multiplicity $N$ (middle) and leading-jet pseudorapidity $\eta^{(1)}$ (right)
for $e \gamma \to e X$. We compare different cuts to suppress the background
from $e^+e^- \to W^\ast_{\to l \bar{\nu}} W^\ast_{\to jj}$,
cutting on either $\eta^{(1)} > 0$, $p_{T,\mathrm{miss}} < 35$ GeV or $N = 1$. }\label{fig:comparison-cut}
\end{figure*}

Based on these observations, we suggest
three different cuts, which aim
at improving the
signal-to-background ratio for a study of the
partonic content of the photon from $e$+hadrons final states.
These are either of the following
\begin{equation}
    N_\mathrm{jets} \le 1 \quad \mathrm{or} \quad
    \eta^{(1)} > 0 \quad \mathrm{or} \quad
    p_{T,\mathrm{miss}} < 35 \ \mathrm{GeV}
\end{equation}
We compare the effect of
these cuts in Fig.~\ref{fig:comparison-cut}
in the respective distributions and observe them to
be largely
orthogonal to each other, i.e.\ cutting out completely separate parts of the phase space.

\begin{figure*}[t]
\centering
\begin{tabular}{cc}
    \includegraphics[width=0.49\linewidth]{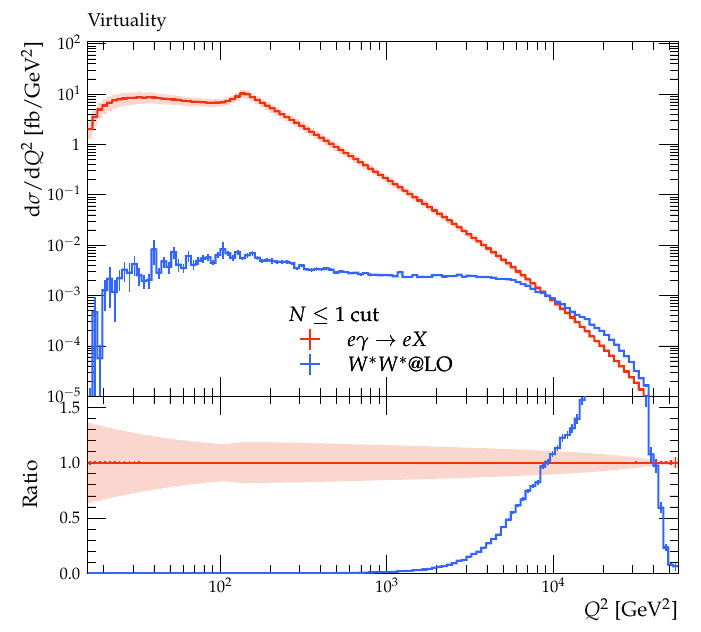} &
    \includegraphics[width=0.49\linewidth]{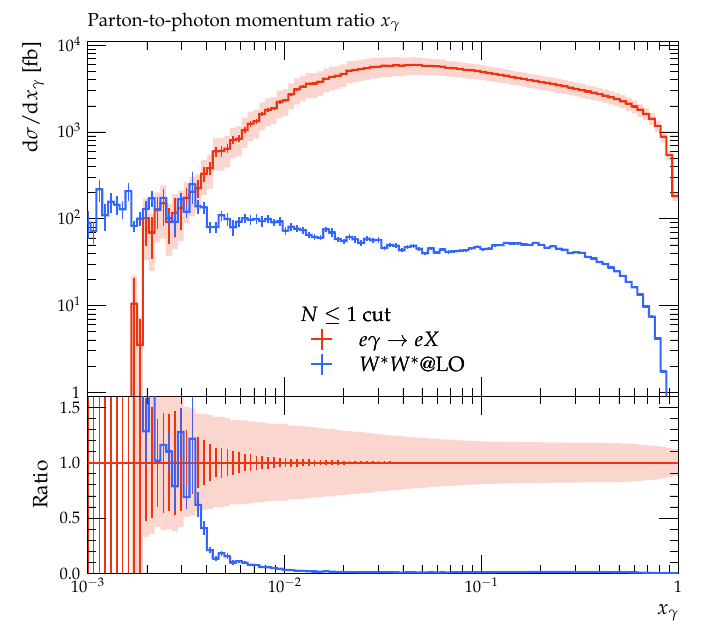} \\
    \includegraphics[width=0.49\linewidth]{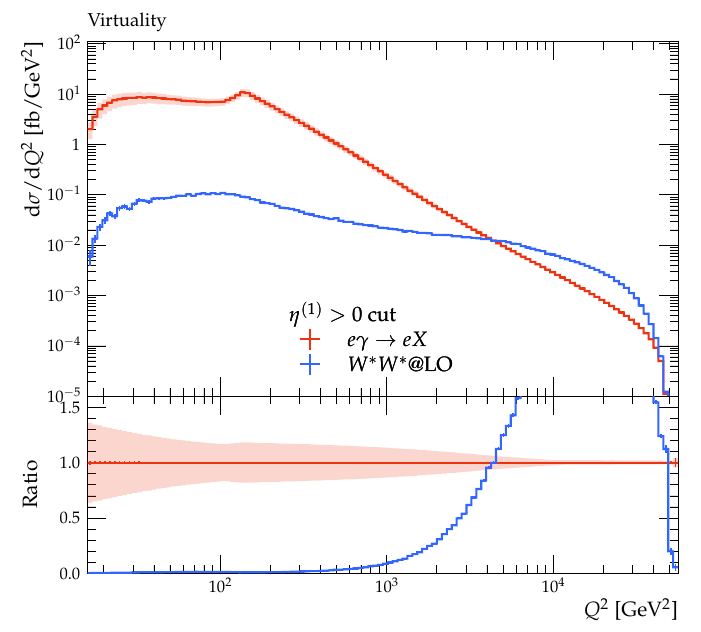} &
    \includegraphics[width=0.49\linewidth]{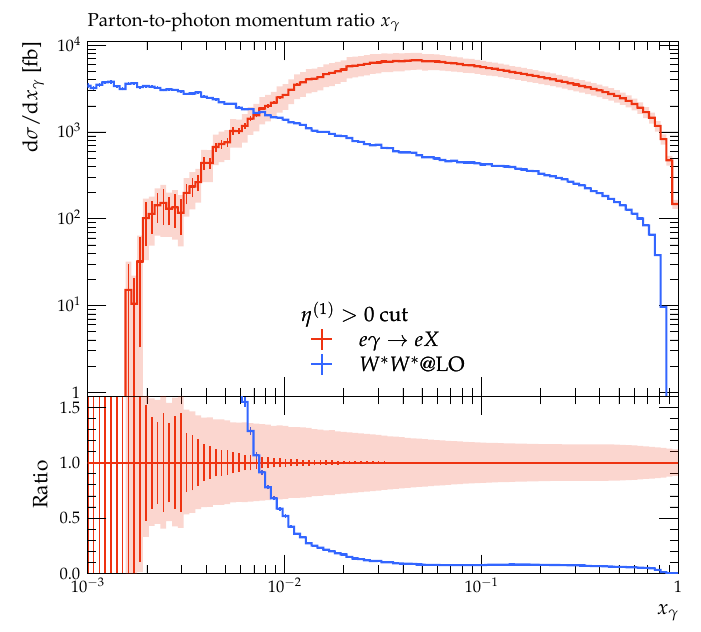} \\
    \includegraphics[width=0.49\linewidth]{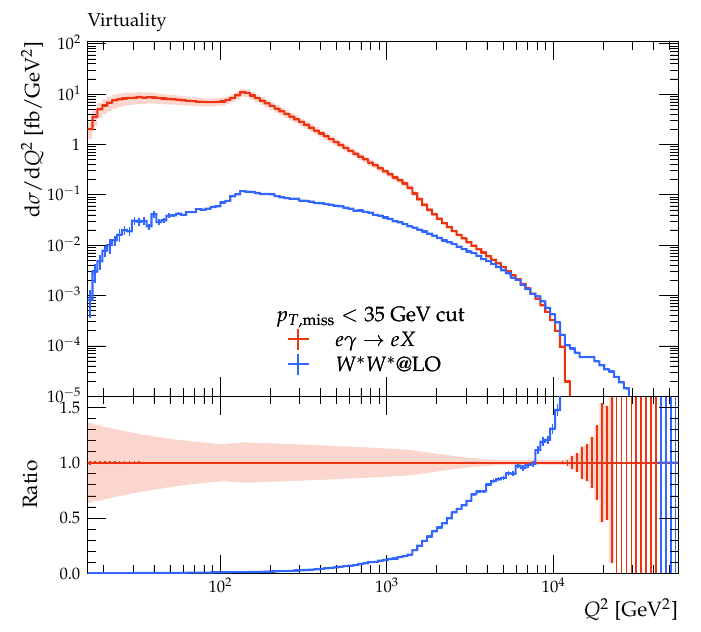} &
    \includegraphics[width=0.49\linewidth]{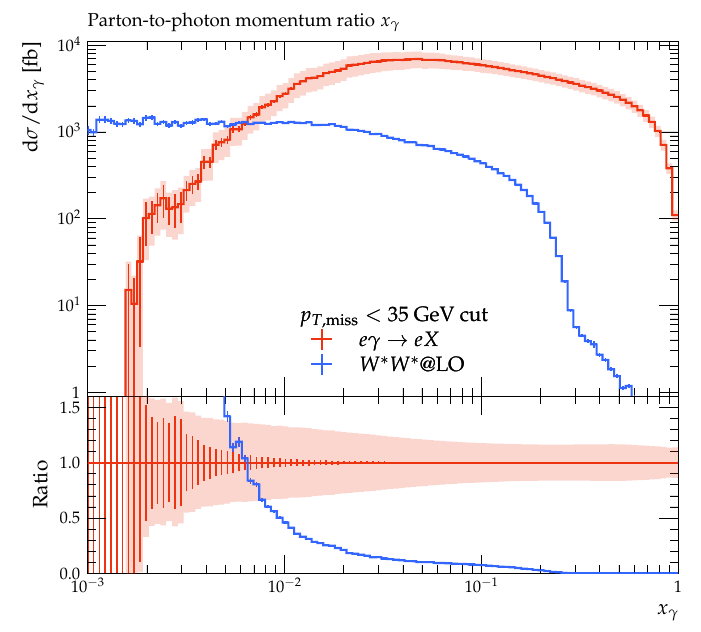}
\end{tabular}
\caption{Distributions of virtuality $Q^2$ (left column) and Bjorken variable $x_\gamma$ (right column)
for $e \gamma \to e X$ and $e^+e^- \to W^\ast_{\to l \bar{\nu}} W^\ast_{\to jj}$
with an additional cut on jet multiplicity $N = 1$ (top row),
leading-jet pseudorapidity $\eta^{(1)} > 0$ (middle row) or
missing transverse momentum $p_{T,\mathrm{miss}} < 35$ GeV (bottom row).}\label{fig:background-with-cuts}
\end{figure*}

The effect of the different cuts on the distributions in $Q^2$ and in $x_\gamma$
is illustrated in Fig.~\ref{fig:background-with-cuts} for the
signal and background processes.
It can be seen that the cut on the jet multiplicity achieves the best enhancement of the
lepton-photon scattering over the $W$-pair background, however, even in that case
the latter dominates for high $Q^2$.
The cuts on leading jet pseudorapidity and
$p_{T,\mathrm{miss}}$ are less performant
in the regions of large $Q^2$, with the latter cut
leading even to a complete suppression of the signal in this
kinematical endpoint region.
Similarly, the $N \le 1$ cut achieves the best reduction of the boson-pair background
for small values of $x_\gamma$, extending the range of a signal-to-background ratio
above unity towards a value of $4 \cdot 10^{-3}$.
As intended, neither of the cuts has
significant impact in the region of small $Q^2$ and
they lead only to a small reduction in the reach of small $x_\gamma$.
As none of the cuts is able to efficiently
suppress the gauge-boson pair background in the large-$Q^2$ region,
which is furthermore affected by large contributions from the direct channels,
this region will be difficult to fully exploit for photon PDF fits.

\section{Conclusion}
\label{sec:conc}
In this study, we considered jet production through lepton-photon scattering
at a future lepton collider. In high-energy collisions the photon
exhibits an intricate partonic structure described by
parton distribution functions.
Measurements of lepton-photon scattering, in analogy to deep inelastic scattering,
can be considered the signal processes to determine these distributions.

We indeed found that jet production at
moderate momentum transfer $Q^2$
is particularly sensitive to the so-called resolved channels,
while the direct channels contribute dominantly at large $Q^2$.
For small values of $Q^2$,
the kinematic reach is mainly limited by the
forward acceptance of the scattered lepton,
which removes substantial parts of the phase space where the cross-section from the resolved channels is large, thus motivating an enhanced forward-lepton coverage for precision
studies of the photon structure.
To underline the constraining power, we compared different PDF parametrisations and saw significant differences.

In studies of lepton-proton scattering, the Breit frame is commonly used for jet studies.
We observed that this is not suitable if a probe of resolved
photon processes is aimed for.
This is a consequence of the fact that in the Breit frame the Born-level leading jet
is boosted into the current direction, hence impossible to be clustered.
The direct channels are  dominant
for higher multiplicities 
and are thus enhanced in the Breit frame.
In contrast, this is not the case in the laboratory frame, making it the preferred choice for jet measurements.
In contrast, Breit-frame event shape observables that use 
the direction of the exchanged boson current as reference axis 
offer very much complementary information to jet cross sections. These event shapes measure the QCD activity along the 
current direction, and thus allow to probe the resolved 
photon process. 

The main background to electron-photon scattering into 
jet final states is the production of $W$ boson pairs,
and we assess different cuts to reduce its contribution.
We find that the best signal-to-background ratio is achieved by employing
a cut on the jet multiplicity, $N \le 1$, in the laboratory frame.
However, the region of large $Q^2$ can not be fully separated from the background. In this region, sensitivity 
on the partonic content of the photon was already low due to 
the dominance of the direct process, and is further diminished 
due to the incomplete background separation. 

Our work outlines the main aspects, opportunities and 
limitations of studying the partonic content of the 
photon through DIS-like jet production
at future lepton colliders. 
More refined studies will require a more detailed modelling of 
the final state acceptance, once detector configurations have 
become mature. 

\FloatBarrier

\acknowledgments

PM thanks Stefan H\"oche for technical advice on the implementation.
This work has been supported by the Swiss High Energy Physics
initiative for the FCC (CHEF), with funding provided specifically by
SERI and the University of Zurich, as well as  by
the Swiss National Science Foundation (SNF) under contract 240015.

\bibliographystyle{JHEP}
\bibliography{bibliography}
\end{document}